\input harvmac
\input epsf

\Title{\vbox{\baselineskip12pt
\hbox{KUNS-1849}
\hbox{{\tt hep-ph/0307361}}}}
{\centerline{Proposal for Generic Size of Large Extra Dimensions}}

\centerline{Satoshi Matsuda\footnote{$^\dagger$}{matsuda@phys.h.kyoto-u.ac.jp}}
\medskip
{\it \centerline{Department of Physics, Graduate School of Science, Kyoto University}
\centerline{Yoshida-Nihonmatsu, Sakyo, Kyoto 606-8501, Japan}}
\bigskip
\centerline{and}
\bigskip
\centerline{Shigenori Seki\footnote{$^\ddagger$}{seki@postman.riken.go.jp}}
\medskip
{\it \centerline{Theoretical Physics Laboratory}
\centerline{The Institute of Physical and Chemical Research (RIKEN)}
\centerline{2-1 Hirosawa, Wako 351-0198, Japan}}

\vskip .3in

\centerline{{\bf Abstract}}

In order to resolve the hierarchy problem, large extra dimensions have been 
introduced, and it has been suggested that the size of extra dimensions is
sub-millimeter. 
On the other hand, we assume in this paper that 
the cosmological constant comes from the Casimir energy of extra dimensions 
and estimate the size of extra dimensions in terms of the 
value of the cosmological constant discovered by the recent WMAP observations.
We demonstrate that this size is consistent 
with the one derived from the hierarchy problem
and propose that there may be a generic size of large extra dimensions
when the number of extra dimensions is equal to two.

\lref\AADD{I. Antoniadis, N. Arkani-Hamed, S. Dimopoulos and G. Dvali, ``New Dimensions at a Millimeter to a Fermi and Superstrings at a TeV,'' Phys. Lett. B436 (1998) 257, {\tt hep-ph/9804398}.}
\lref\ADD{N. Arkani-Hamed, S. Dimopoulos and G. Dvali, ``The Hierarchy Problem and New Dimensions at a Millimeter,'' Phys. Lett. B429 (1998) 263, {\tt hep-ph/9803315}.}
\lref\AW{J. Ambj{\o}rn and S. Wolfram, ``Properties of the Vacuum 1. Mechanical and Thermodynamic,'' Annals Phys. 147 (1983) 1; ``Properties of the Vacuum 2. Electrodynamic,'' Annals Phys. 147 (1983) 33.}
\lref\Gu{A. Gupta, ``Contribution of Kaluza-Klein Modes to Vacuum Energy in Models with Large Extra Dimensions \& the Cosmological Constant,'' {\tt hep-th/0210069}.}
\lref\GS{I. M. Gel'fand and G. E. Shilov, ``Generalized Functions,'' Volume 1 (Academic Press, 1964).}
\lref\HLZ{T. Han, J. D. Lykken and R.-J. Zhang, ``Kaluza-Klein States from Large Extra Dimensions,'' Phys. Rev. D59 (1999) 105006, {\tt hep-ph/9811350}.}
\lref\WMAP{C. L. Bennett {\it et al.}, ``First Year Wilkinson Microwave Anisotropy Probe (WMAP) Observations: Preliminary Maps and Basic Results,'' {\tt astro-ph/0302207}; see also the Web page, {\tt http://map.gsfc.nasa.gov/}.}
\lref\LCCGVP{J. C. Long, H. W. Chan, A. B. Churnside, E. A. Gulbis, M. C. M. Varney and J. C. Price, ``New Experimental Limits on Macroscopic Forces Below 100 Microns,'' letter version published in Nature 421 (2003) 922, {\tt hep-ph/0210004}.}
\lref\ADDi{N. Arkani-Hamed, S. Dimopoulos and G. Dvali, ``Large Extra Dimensions: A New Arena for Particle Physics,'' Physics Today 55 (2002) 35.}
\lref\Me{A.C.Melissinos, ``Vacuum Energy and the Cosmological Constant,'' {\tt hep-ph/0112266}.}
\lref\Mi{K. A. Milton, ``Dark Energy as Evidence for Extra Dimensions,'' {\tt hep-ph/0210170}.}
\lref\Ad{E. G. Adelberger for the E{\" o}t-Wash Group, ``Sub-mm Tests of the Gravitational Inverse-square Law,'' {\tt hep-ex/0202008}.}

\Date{July 2003}

\newsec{Introduction}

Large extra dimensions have been introduced in Refs.\refs{\AADD,\ADD} 
in order to resolve the hierarchy problem. 
The interesting possibility of their existence 
has stimulated much theoretical \refs{\Me\Mi{--}\Gu} and 
experimental research \refs{\LCCGVP\ADDi{--}\Ad}.

On the other hand, the origin of the cosmological constant 
has not been made clear so far. 
This is one of the most important problems in physics.
Recently the WMAP project \WMAP\ has determined accurately
several parameters of our universe, 
for example, the age of universe, the density of dark matter and so on.

In section 2 of this paper we calculate the Casimir energy derived 
from extra dimensions. 
In section 3 we assume that the Casimir energy can be identified with 
the cosmological constant, and estimate the size of extra dimensions 
by the use of the observed value of the cosmological constant.
Section 4 is devoted to the conclusions, where we compare 
the size of extra dimensions corresponding to the cosmological constant
problem with the one corresponding to the hierarchy problem.

\newsec{Casimir energy and cosmological constant}

We consider the $d$ dimensional compact extra space
which is the product of $d$ spheres, $(S^1)^{\otimes d}$.
We set that the size of every $S^1$ is $R$.
The Kaluza-Klein modes of gravitons are generated from this space and
contribute to the vacuum energy in our four dimensional space-time.
The vacuum energy is obtained as the Casimir energy derived from
the extra space, and is given by
\eqn\vacene{
E=\left[{(2+d)(3+d) \over 2} -1\right]
{1 \over 2} \left(\prod_{i=1}^d \sum_{n_i=-\infty}^\infty\right)
\int {d^3k \over (2\pi)^3}
\sqrt{{\bf k}^2 + M^2},
}
where $M^2 \equiv 4\pi^2 \sum_{i=1}^d n_i^2/R^2$ 
and the factor in front, $(2+d)(3+d)/2-1$, represents the number of 
polarization degrees of freedom for a graviton in $4+d$ dimensions \HLZ.
Since this integration is generally divergent,
we use dimensional regularization and Eq.\vacene\ becomes
\eqn\regene{
E = -{1 \over 64\pi^2} \left[{(2+d)(3+d) \over 2} -1\right]
\left(\prod_{i=1}^d \sum_{n_i=-\infty}^\infty\right)
M^4 \left[\ln \left({\lambda^2 \over M^2}\right) + {3 \over 2}\right].
}
$\lambda$ is a parameter with mass dimension.
We calculate the following infinite summations,
\eqnn\infsum
$$\eqalignno{
\left(\prod_{i=1}^d \sum_{n_i=-\infty}^\infty\right)
\left( \sum_{i=1}^d n_i^2 \right)^2
&= d \left(\prod_{i=1}^d \sum_{n_i=-\infty}^\infty\right) n_1^4
+ d(d-1) \left(\prod_{i=1}^d \sum_{n_i=-\infty}^\infty\right) n_1^2 n_2^2 \cr
&= 2d \zeta(0)^{d-1} \zeta(-4) + 4d(d-1) \zeta(0)^{d-2} \zeta(-2)^2 \cr
&= 0, &\infsum
}$$
where we have used the zeta function equalities $\zeta(-4) = \zeta(-2) = 0$.
Using Eq.\infsum, the vacuum energy \regene\ is reduced to
\eqn\regenei{
E = {\pi^2 \over 2R^4} \left[{(2+d)(3+d) \over 2} -1\right]
\left(\prod_{i=1}^d \sum_{n_i=-\infty}^\infty\right)
{d \over dx} N^x \bigg|_{x=4},
}
where $N \equiv \sqrt{\sum_{i=1}^d n_i^2}$.
In order to take care of a divergence of the right hand side of Eq.\regenei,
we have to regularize it.
Since it is difficult to calculate it analytically,\footnote*{In the
special cases of $d$, we can obtain exact results in terms
of Epstein zeta functions \AW.}
let us consider a numerical method.
In terms of the Fourier transformation \GS, $N^x$ is described as
\eqn\fourier{
N^x = -{2^x \sin {\pi x \over 2} \over \pi^{{d \over 2}+1}}
\Gamma\left(1+{x \over 2}\right) \Gamma\left({x+d \over 2}\right)
\int q^{-x-d} e^{i {\bf N}\cdot{\bf q}} d{\bf q},
}
where ${\bf N} \equiv (n_1,\cdots,n_d)$, ${\bf q} \equiv (q_1,\cdots,q_d)$
and $q \equiv |{\bf q}|$, while the infinite summation of
$\exp (i {\bf N}\cdot{\bf q})$ is expressed by
\eqn\sumdelta{
\left(\prod_{i=1}^d \sum_{n_i=-\infty}^\infty\right)
e^{i {\bf N} \cdot {\bf q}}
= (2\pi)^d
\prod_{i=1}^d \left[\sum_{m_i=-\infty}^\infty \delta(q_i - 2\pi m_i)\right].
}
From Eqs.\fourier\ and \sumdelta, we obtain
\eqn\nsum{
\left(\prod_{i=1}^d \sum_{n_i=-\infty}^\infty\right) N^x
= -{\sin{\pi x \over 2} \over \pi^{1+x+{d \over 2}}}
\Gamma\left(1+{x \over 2}\right)\Gamma\left({x+d \over 2}\right)
\left(\prod_{{i=1 \atop {\bf m} \neq {\bf 0}}}^d
\sum_{m_i=-\infty}^\infty\right)
\left(\sum_{i=0}^d m_i^2\right)^{-{x+d \over 2}},
}
where ${\bf m}\equiv (m_1,\cdots,m_d)$.
Using this equation, the vacuum energy \regenei\ becomes
\eqnn\energy
\eqnn\msum
$$\eqalignno{
E &= - {1 \over 2 \pi^{2+{d \over 2}} R^4} \left[{(2+d)(3+d) \over 2} -1\right]
\Gamma\left(2+{d \over 2}\right) {\cal M}_d, &\energy\cr
&{\cal M}_d \equiv
\left(\prod_{{i=1 \atop {\bf m} \neq {\bf 0}}}^d
\sum_{m_i=-\infty}^\infty\right)
\left(\sum_{i=0}^d m_i^2\right)^{-2-{d \over 2}}. &\msum
}$$
The divergence coming from the infinite summations in Eq.\regenei\
is thus resolved and Eq.\energy\ converges.
Calculating Eq.\msum\ numerically, we obtain
\eqn\msumres{
{\cal M}_1 \approx 1.03693,\quad
{\cal M}_2 \approx 4.6589,\quad
{\cal M}_3 \approx 7.46706.
}
Substituting \msumres\ into Eq.\energy, 
the vacuum energies for $d=1,2,3$ become
\eqn\periodic{
E \approx {1 \over R^4}\cases{
-0.196993,&\quad $d=1$, \cr
-1.35231, &\quad $d=2$, \cr
-3.16082, &\quad $d=3$.
}}
Note that the vacuum energies corresponding to the higher dimensions 
are similarly calculable.
The result for every dimension turns out to be negative.

In the following sections 
we will suggest that the vacuum energies correspond 
to the cosmological constant. 
Since the energies in \periodic\ are negative, we can not 
adopt such identification. 
So far we have considered the extra space which is $(S^1)^{\otimes d}$,
that is, the $d$ dimensional compact space with a periodic boundary condition.
On the other hand, if the extra space has a Dirichlet boundary condition, 
the vacuum energies may become positive.
Let us consider the $d$ dimensional compact extra space with 
the Dirichlet boundary condition in the rest of this section.

Since the wave function becomes a sine function on account of 
the Dirichlet boundary condition, 
the vacuum energy is modified from Eq.\vacene\ to the form
\eqn\dirvacene{
E={1 \over 2} \left[{(2+d)(3+d) \over 2} -1\right]
\left(\prod_{i=1}^d\sum_{n_i=1}^\infty\right)
\int {d^3k \over (2\pi)^3}
\sqrt{{\bf k}^2 + M^2}.
}
We rewrite this equation as
$$\eqalign{
E&={1 \over 2} \left[{(2+d)(3+d) \over 2} -1\right]
\prod_{i=1}^d 
\left[{1 \over 2}\left(\sum_{n_i=-\infty}^{-1}
+\sum_{n_i=1}^\infty \right)\right]
\int {d^3k \over (2\pi)^3} \sqrt{{\bf k}^2 + M^2} \cr
&={1 \over 2 \cdot 2^d} \left[{(2+d)(3+d) \over 2} -1\right]
\prod_{i=1}^d \left[\sum_{n_i=-\infty}^\infty (1 - \delta_{n_i 0})\right]
\int {d^3k \over (2\pi)^3} \sqrt{{\bf k}^2 + M^2} \cr
&= {\pi^2 \over 2^{d+1} R^4} \left[{(2+d)(3+d) \over 2} -1\right]
\prod_{i=1}^d \left[\sum_{n_i=-\infty}^\infty (1 - \delta_{n_i 0})\right]
{d \over dx} N^x \bigg|_{x=4}.
}$$
After numerical calculations, we obtain
\eqn\dirichlet{
E \approx {1 \over R^4} \cases{
-0.196993,  &\quad $d=1$,\cr
0.0165101,  &\quad $d=2$,\cr
-0.0199402,  &\quad $d=3$, \cr
0.00149895,&\quad $d=4$.
}}
In the results \dirichlet, when the numbers of extra dimensions are even, 
the vacuum energies become positive.

\newsec{The cosmological constant and the size of extra dimensions}

Recently the WMAP project has determined the several 
parameters of our universe \WMAP.  
Let us consider the cosmological constant by using the data 
given by WMAP.
Since the Hubble constant is
\eqnn\hubble
$$\eqalignno{
H_0 &= h \times 100 \ [{\rm km} \cdot {\rm sec}^{-1} \cdot {\rm Mpc}^{-1}] \cr
  &= 0.7675 \times 10^{-28} \ [{\rm cm}^{-1}], &\hubble
}$$
where $h=0.71$ (according to the WMAP observation), 
the critical density of universe becomes
\eqnn\cridense
$$\eqalignno{
\rho_c &= {3 H_0^2 \over 8 \pi G} \cr
       &= 5.313 \times 10^3 \ [{\rm eV} \cdot {\rm cm}^{-3}], &\cridense
}$$
where $G$ is the gravitation constant.
WMAP gives for the total density $\Omega_{\rm tot}$ 
and the dark energy density $\Omega_\Lambda$,
\eqn\wmapdata{\eqalign{
\Omega_{\rm tot} &= 1.02 \pm 0.02,\cr
\Omega_\Lambda &= 0.73 \pm 0.04.
}}
Note that the dark energy is the same as the cosmological constant.
In terms of the data \cridense\ and \wmapdata, we calculate 
the cosmological constant to be
\eqnn\cosmoconst
$$\eqalignno{
\Lambda &= \rho_c \Omega_{\rm tot} \Omega_\Lambda \cr
        &= 3.956 \times 10^3 \ [{\rm eV} \cdot {\rm cm}^{-3}]. &\cosmoconst
}$$
We suggest that the cosmological constant is identified 
with the vacuum energy derived from the extra dimensions.

In the previous section, we have obtained the vacuum energy 
with the positive values when the number of 
extra dimensions is even and the extra spaces have the Dirichlet 
boundary conditions. 
Let us transform the canonical values \dirichlet\ of the vacuum energy 
in units of $\hbar c = 1.973269 \times 10^{-5} \ [{\rm eV} \cdot {\rm cm}]$ 
to the physical values. 
Then the vacuum energy is calculated to be 
\eqn\canodir{
E \ [{\rm eV} \cdot {\rm cm}^{-3}] ={1 \over R^4}
\cases{
3.25787 \times 10^{-7},\quad &$d=2$, \cr
2.95782 \times 10^{-8},\quad &$d=4$,
}}
where the unit of $R$ is centimeter.
If we identify the cosmological constant \cosmoconst\ with 
the vacuum energy \canodir, 
the size $R$ of extra dimensions then becomes
\eqn\extsize{
R \ [{\rm cm}] = 
\cases{
0.003012,\quad &$d=2$, \cr
0.001654,\quad &$d=4$.
}}

\newsec{Conclusions}

In Ref.\ADD, the scale of extra dimensions is discussed from 
the viewpoint of the hierarchy problem. 
The four dimensional Planck scale $M_{\rm pl}$ is related to 
the $4+d$ dimensional one $M_{{\rm pl}(4+d)}$by 
$M_{\rm pl}^2 \sim M_{{\rm pl}(4+d)}^{2+d} R^d$.
If we put $M_{{\rm pl}(4+d)} \sim m_{\rm EW} \sim 1 \ [{\rm TeV}]$, 
where $m_{\rm EW}$ is an electro-weak scale, 
the size of extra dimensions then becomes 
\eqn\extsizehie{
R \sim 10^{{30 \over d} - 17} \ [{\rm cm}].
}
So $R$ has a sub-millimeter size of $10^{-2} \ [{\rm cm}]$ 
when the number $d$ of extra dimensions is equal to two.
If $d=4$, $R$ has a size of $10^{-10} \ [{\rm cm}]$.

On the other hand, we have adopted the suggestion 
that the cosmological constant can be 
identified with the vacuum energy which is 
 the Casimir energy coming from the space of extra dimensions.
By this identification we have calculated the sizes of extra dimensions.
We have used the recent results given by the WMAP project.
In \extsize, when the number $d$ of extra dimensions equals 
two, the size $R$ of extra space becomes $3.012 \times 10^{-3} \ [{\rm cm}]$. 
When $d=4$, the size $R$ is $1.654 \times 10^{-3} \ [{\rm cm}]$.

Comparing these sizes of extra dimensions obtained from the 
two independent points of view, that is, 
the hierarchy problem and the cosmological 
constant, the two sizes for $d=2$ turn out to be almost consistent. 
So we propose that there may be the generic size 
of large extra dimensions for $d=2$, which is 
in the range of $10^{-2}${--}$10^{-3} \ [{\rm cm}]$.

Finally let us conclude with a brief comment that our proposal 
also resolves the critical issue often raised where the enormous
gap of energy scale in the hierarchy problem is just transmuted into an 
arbitrarily chosen size of large extra dimensions.

\bigbreak\bigskip\bigskip\centerline{{\bf Acknowledgement}}\nobreak
S. M. thanks Theoretical Physics Group of RIKEN for providing 
a pleasant stay for him to finish this paper.
S. S. is grateful to M. Sakamoto for useful lectures on WMAP.

\listrefs
\bye